\documentclass[11pt]{article}
\usepackage{times}
\usepackage{geometry}
\usepackage[utf8]{inputenc}
\usepackage{enumitem,amssymb}
\usepackage{graphicx}
\usepackage{fancyhdr}
\usepackage{aas_macros}
\usepackage{mdframed} 

\usepackage[authoryear]{natbib}
\setcitestyle{authoryear,open={(},close={)}}
\mdfdefinestyle{theoremstyle}{innertopmargin=\topskip}
\mdtheorem[style=theoremstyle]{lrptextbox}{}

\begin{document}

\section*{Machine Learning Advantages in Canadian Astrophysics}

The application of machine learning (ML) methods to the analysis of astrophysical datasets is on the rise, particularly as the computing power and complex algorithms become more powerful and accessible. As the field of ML enjoys a continuous stream of breakthroughs, its applications demonstrate the great potential of ML, ranging from achieving tens of millions of times increase in analysis speed (e.g., modeling of gravitational lenses or analysing spectroscopic surveys) to solutions of previously unsolved problems (e.g., foreground subtraction or efficient telescope operations). The number of astronomical publications that include ML has been steadily increasing since 2010.
 With the advent of extremely large datasets from a new generation of surveys in the 2020s, ML methods will become an indispensable tool in astrophysics.  Canada is an unambiguous world leader in the development of the field of machine learning, attracting large investments and skilled researchers to its prestigious AI Research Institutions. This provides a unique opportunity for Canada to also be a world leader in the application of machine learning in the field of astrophysics, and foster the training of a new generation of highly skilled researchers. 



\section*{Authors}

{\bf Kim Venn} (University of Victoria), 
{\bf S\'ebastien Fabbro} (NRC-Herzberg),
{\bf Adrian Liu} (McGill), 
{\bf Yashar Hezaveh} (Universit\'e de Montr\'eal), 
{\bf Laurence Perreault-Levasseur} (Universit\'e de Montr\'eal, MILA), 
{\bf Gwendolyn Eadie} (University of Toronto), 
{\bf Sara Ellison} (University of Victoria), 
{\bf Joanna Woo} (Simon Fraser University), 
{\bf JJ Kavelaars} (NRC-Herzberg), 
{\bf Kwang Moo Yi} (University of Victoria), 
{\bf Ren\'ee Hlo\~zek} (University of Toronto), 
{\bf Jo Bovy} (University of Toronto),
{\bf Hossen Teimoorinia} (NRC-Herzberg), 
{\bf Siamak Ravanbakhsh} (McGill, MILA),
{\bf Locke Spencer} (University of Lethbridge)

\section{Introduction}

Applying machine learning (ML) methods to analyzing astrophysical datasets have become extremely popular,
particularly as the computing power and complex algorithms become more powerful and accessible.
Large observational surveys, as well as simulations, have provided massive datasets for developing ML tools with astrophysical applications, making ML ever more tempting. 
As the field of ML enjoys a continuous stream of breakthroughs, its applications in astronomy demonstrate its great potential.
These examples include achieving tens of millions of times increase in analysis speed, for example in modeling gravitational lenses or analysing spectroscopic surveys, as well as introducing new solutions to previously unsolved problems, such as foreground subtraction or efficient telescope operations.
With the advent of extremely large datasets from a new generation of surveys in the 2020s, ML methods will become an indispensable tool in astrophysics.

\section{The Rise in Machine Learning}

Today’s widespread use of artificial intelligence (AI), and more specifically ML, a certain class of approaches to AI, can be traced back to the beginning of the ‘deep learning revolution’, when the 2012 {\it ImageNet} challenge was won by Krizhevsky et al.
Originally published in 2009, {\it ImageNet} is an image database containing more than 14 million hand-annotated images.
Since 2010, the ImageNet project runs an annual competition called the ImageNet Large Scale Visual Recognition Challenge (ILSVRC), where algorithms compete to identify objects present in a subset of images from the dataset.
In 2012, a group from the University of Toronto submitted a deep convolutional neural network (CNN) architecture called {\it AlexNet} (which is still used in research to this day) that outperformed previous models and nearest contenders by a margin of more than 10\% in accuracy.
One of the strengths of this model when compared to its brittle competitors, and which enabled its widespread use across many fields, is the data-driven nature of its training: while the training was performed on a specific dataset, the architecture of the model itself is completely general and could, in principle, be used to perform learning on practically any visual dataset.

\begin{figure}
    \centering
    \includegraphics[scale=0.3]{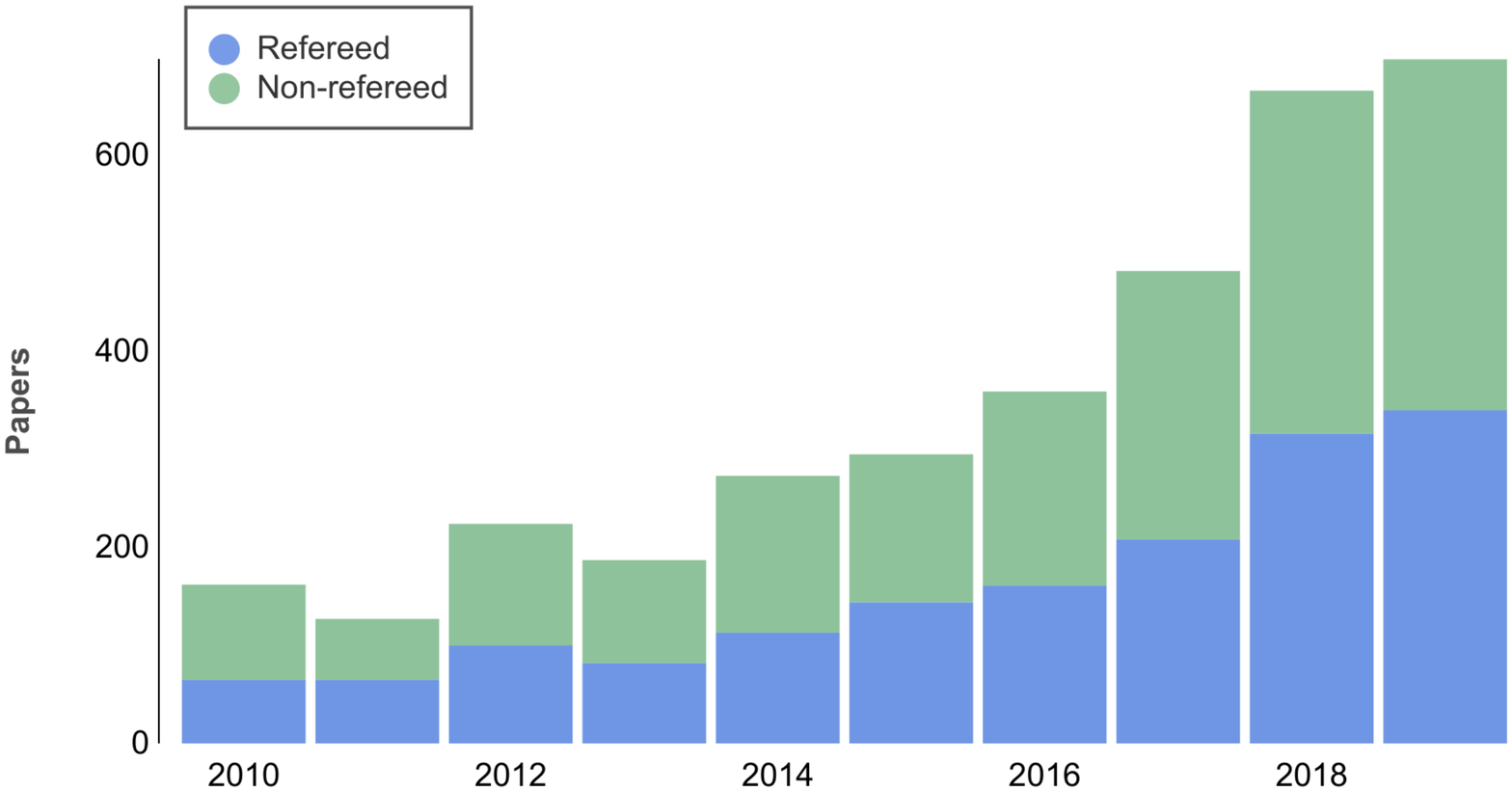}
    \caption{Astronomy papers that include machine learning methods in the abstract or title since 2010.   From the Astrophysics Data System (ADS).}
    \label{fig:mlpapers}
\end{figure}

By 2014, all the high-scoring {\it ImageNet} competitors were using deep architectures. 
Since then, other high-profile datasets have been introduced for deep learning by Google, Microsoft, and the Canadian Institute for Advanced Research\footnote{ https://qz.com/1034972/the-data-that-changed-the-direction-of-ai-research-and-possibly-the-world/}.
Today, these convolutional neural networks are omnipresent beyond the confined world of academic challenges, including but not limited to, {\it Facebook}, self-driving cars, natural language processing, fraud detection, robot navigation, medical diagnostics, targeted marketing, and gaming.
They are readily applicable to anything requiring image, audio or video processing.  



CNNs and many deep learning methods learn patterns between nearby pixels on ascending levels of abstraction to produce outputs of interest, using thousands to millions of computations at each level.
In CNNs, at each of these levels, referred to as {\it layers},  this is done by processing the image by convolving it with a number of filters. 
The resulting maps are then fed to the following layer as an input. 
After a number of these layers, the output of the last layer is interpreted as the output of the network. The values of the filters, also called network weights, are learned through a process known as {\it training}, where pairs of correct input-output examples are shown to the network. Given enough training examples, these networks can make accurate predictions on previously unseen examples using these learned parameters.

While CNNs date back to well before the beginning of the 21st century, the keys to their recent success -- proper initialization~\citep{Glorot2010}, advanced activation functions~\citep{Nair2010}, better solvers~\citep{Kingma2015}, and the use of Graphic Processing Units (GPU) for high performance computing -- were not developed until the last decade.
Therefore, the 1990's saw a number of attempts at using neural networks for, e.g., spectral classifications and parameter estimation\citep[e.g.,][]{bailerjones1998, Gulati1994} and for star/galaxy separation through algorithms such as SExtractor \citep{SEXtractor1996},
but they were not successful.
Deep learning only started to gain popularity among the astronomy community around 2016, after the CASCA Mid-Term Review of LRP2010. In Fig.~\ref{fig:mlpapers}, the rise in ML applications in astronomy can be seen to have sky rocketed, now reaching about 2 papers/day. 
 
Even with this recent growth in popularity, astronomy still represents a golden, mostly unexplored opportunity for machine learning, given the existence of relatively large, homogeneous datasets\footnote{
Deep learning is known to benefit immensely from data (the `` unreasonable effectiveness of data''), as 
demonstrated by \cite{Sun2017}.}
for a range of applications from imaging to spectroscopy.
The Sloan Digital Sky Survey (SDSS) DR15\footnote{https://www.sdss.org/dr15/data\_access/volume/}
alone provides over 170 TB data (of this $\sim50$\% is raw or intermediate data), where $>70$ TB is APOGEE spectra (raw, reduced, or synthetic), $>60$ TB is eBOSS photometric data (raw or reduced), $>20$ TB is eBOSS spectroscopic data, and $>13$ TB is MaNGA spectra (raw or reduced 2d individual exposures and 3d summary stacks).  The datasets associated with the Large Synoptic Survey Telescope (LSST) and the Square Kilometre Array (SKA) are expected to dwarf these, e.g., LSST is expected to obtain 20 TB per night, and SKA is estimated to run at 160 TB per second.  These latter data rates are too high to record all of the raw data, which is much like the LHC at $\sim600$ TB per second\footnote{https://home.cern/science/computing/processing-what-record}, forcing astronomers to automate the selection of only the most interesting events to record.

 One of the early successful applications of deep learning in astronomy was in the field of strong gravitational lensing, where CNNs were used for performing both the tasks of lens finding \citep[e.g.][]{Lanusse2017,Jacobs:17,Petrillo:18,Pourrahmani:18,Schaefer:18}, and lens modeling \citep{Hezaveh2017,Perreault2017,Morningstar2018, Morningstar2019}, automating and accelerating the inference of lens parameters by many orders of magnitude (about ten million times faster on a single graphics processing unit), without loss of accuracy when compared to time- and resource-consuming traditional methods.  
 In the coming years, hundred of thousands of new gravitational lenses from large surveys (e.g., Euclid, LSST), existing and new facilities (e.g., ALMA, JWST, TMT), will provide an opportunity to transform this field. The further development of these promising machine learning-based analysis methods and their implementation in analysis pipelines will allow us to fully exploit the wealth of these upcoming data and circumvent difficulties faced by traditional maximum likelihood methods, 
 allowing us, for example, to map the distribution of matter on small-scales to high precision, opening a new window for testing dark matter models.

\begin{figure}
    \centering
    \includegraphics[scale=0.4]{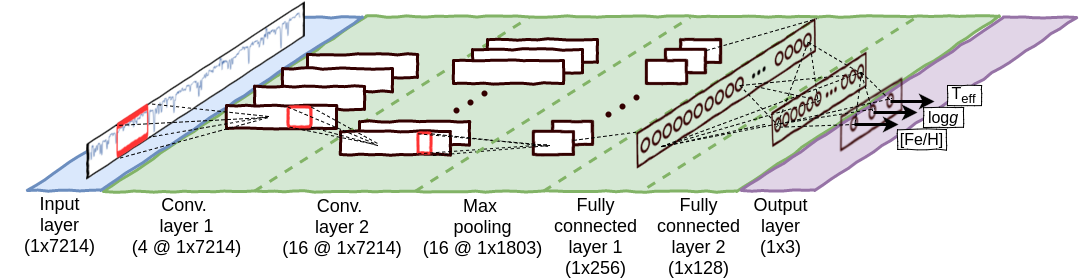}
    \caption{The CNN used by StarNet to analyse SDSS-APOGEE spectra \citep[see][]{fabbro2018}.}
    \label{fig:CNN}
\end{figure}

 An additional early success of ML in astrophysics has been the use of a deep neural network architecture to analyse both observed and synthetic stellar spectra (see Fig.~\ref{fig:CNN}).  \citet{fabbro2018} showed that the stellar parameters (temperature, gravity, and metallicity) from entire SDSS-III APOGEE spectral database can be determined with similar precision and accuracy as the APOGEE pipeline in only a few seconds with machine learning.  The data-driven ML model was further developed by \citet{leung2019} to compute  chemical abundances for over 15 elements with higher precision than the APOGEE (or other) data reduction pipelines. The data-driven ML model was also modified to analyse nearly 1 million spectra from LAMOST \citep{zhang2019} with excellent performance and significantly improved precision.
 In the upcoming era of spectroscopic surveys, ML will be invaluable for fast, efficient, and precise analyses.
 
We summarize a few more important and successful applications of ML in the astronomical literature:
\begin{itemize}[noitemsep,topsep=0pt]
    \item The detection of complex, rare, or new structures in surveys, such as gravitational lenses \citep{Lanusse2018}, supernovae \citep{Moss18}, or galaxy mergers \citep{Ackermann2018}, etc;
    \item Surrogate modelling of computationally expensive simulations, where ML models can act as extremely fast interpolators or emulators, for example in planetary atmosphere simulations \citep{Zingales2018}, in the production of otherwise extremely expensive hydrodynamical simulation to produce HI maps \citep{Fernandez:19}, or in the
    extrapolation of large scale cosmological structure formation
    \citep{Yin2019};
    \item The morphological classifications of galaxies \citep{Dieleman2015}; 
  
    \item The removal of systematic contaminants in data, for example in terrestrial radio frequency interference in radio telescope data \citep{Kerrigan2019}, in foreground removal (such as dust) in CMB data \citep{Aylor:19} and in intensity mapping data, or in cases of difficult background removal (RFI in radio data, cosmic rays in space, bright star halos in wide fields)
    
    \item{
The Photometric LSST Astronomical Time Series Classification Challenge \citep{plasticc}, an open ML challenge hosted on Kaggle\footnote{https://www.kaggle.com/c/PLAsTiCC-2018} with over 1000 teams participating in the challenge, many of whom where not astronomers. This challenge illustrated the power of astronomical data to engage diverse groups interested in machine learning techniques and methodologies. }
    \item{Other examples related to time domain astronomy challenges are reviewed by \cite{hlozek2019}.
    }
\end{itemize}

\begin{figure}
    \centering
    \includegraphics[scale=0.3]{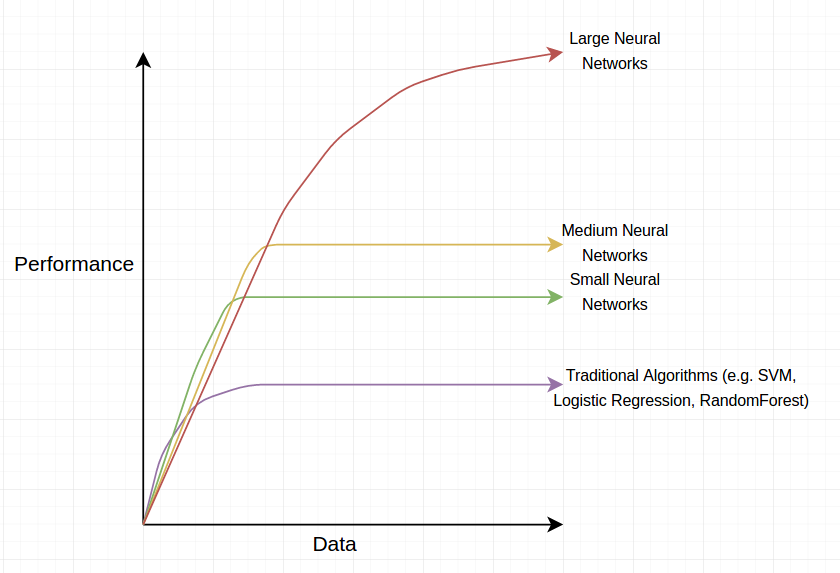}
    \caption{Comparison of ML techniques.   Near the origin, where the size of the dataset is small, different representations are not that distinct, and improvements in feature engineering perform better. As data sizes increase, deep neural networks are able to benefit from their over-parameterized redundancy and capture representations better. Thus, there is a data volume and variety requirement for selecting the best alogrithms foran application.}
    \label{fig:Ng}
\end{figure}

\section{The next decade for ML in Astronomy}

In the coming decade, several astronomy programs will provide very large and homogeneous dataset that are ideal for ML applications.
It is a challenge to consider how we are going to access and analyse such large datasets given that the data volume will increase more rapidly than the network speed; e.g., see Fig.~\ref{fig:Ng}. 
ML methods benefit greatly from larger data sets, and will require access to storage and computing resources in order to exploit fully the potential of multi-surveys. Machine learning infrastructure platforms, with fast data access, easy workflow management, and complex software pipeline deployments have been key for the  success of the AI in industry (e.g., see statements from The State of AI 2019\footnote{https://www.stateof.ai} and AI for the Real World\footnote{https://hbr.org/2018/01/artificial-intelligence-for-the-real-world}). However, for astronomy research, the access to the resources and computing infrastructure has not kept up with the fast pace development observed in industry data science platforms. 
At minimum, establishing a more robustly resources CANFAR science portal is required to enable this.
Tight cooperation with national research infrastructure, but also the new AI Research Institutes (next Section) will be necessary to take full advantage of the revolutions coming from AI research. 

This leap will be particularly important in the next decade with very large scale surveys.  

\begin{itemize}[noitemsep,topsep=0pt]

\item Euclid will be launched by ESA in 2022 and will observe 15,000 deg$^2$ of the darkest sky that is free of contamination by light from our Galaxy and our Solar System. About 10 billion sources will be observed, where $\sim$1 billion will be used for weak lensing and several tens of millions for galaxy redshifts to be used for galaxy clustering, the evolution of cosmic structures, and nature of dark matter.  The complete survey will represent hundreds of thousands of images and several tens of Petabytes of data.  

\item The LSST will begin in 2020 in Chile with an 8.4-meter telescope and 3-gigapixel camera to produce a wide-field astronomical survey of the universe.  It will photograph the entire available sky every few nights, collecting 20 TB of raw image data every night, and processed in near-real time to produce alert notifications of new and unexpected astronomical events. The data will be reprocessed annually to create a 500 PB data archive by the end of the 10-year mission.  Canada has ambitions to join the LSST project by constructing a 20 PB public imaging and catalog archive as in-kind contribution to the project.  This science archive will be coupled with the Canadian LSST Alert Science Platform (CLASP).  CLASP will supply computing hardware and the platform interface needed to make optimal use of the LSST Alert Stream.

\item The Canadian HI Observatory and Radio transient Detector (CHORD) is a proposed next-generation radio interferometer, which along with the SKA, are examples of large radio instruments to be expected in the next decade. 
Both instruments will combine signals received from hundreds to thousands of antennas spread over several thousand kilometres.
Already, similar projects such as the Canadian Hydrogen Intensity Mapping Experiment (CHIME) and the Hydrogen of Epoch of Reionization Array (HERA) require custom hardware for on-site processing, with data flows on the order of 13 terabits of data/second---more than all Canada’s internet traffic. Even after compression, these telescopes have data rates of order $\sim 50TB$ per day. The anticipated data rates will only get more extreme with CHORD and the SKA, posing huge problems in both storage and analysis that will necessitate even better algorithms for real-time signal processing, compression, and extraction of signals of interest. Some estimates for the SKA, for example, suggest that the array could generate an exabyte/day of raw data, possibly compressed to $\sim$ 10 PB/day.

\item Multi-object spectroscopic facilities (e.g., the ESO-4MOST, INT-WEAVE, Subaru-PFS, SDSS-V, and the Canadian-led Maunakea Spectroscopic Explorer) are planned for the 2020's to obtain high quality spectra of several thousand objects simultaneously.  Spectra need to be converted into physical parameters to be used to reveal the structure and dynamics of the Milky Way galaxy, the nature of dark matter and dark energy, the formation and evolution of galaxies, and the structure of the cosmos.  These surveys will deliver over $>25$ million spectra {\it each} over the whole sky, requiring intensive data analysis, including time domain information.   

\end{itemize}

\section{Opportunities in Canada}

 Big data has become an essential tool for scientific progress, underpinning world-class research across all disciplines, including astronomy. 

{\bf Canadian Government Support:} 
Through Budget 2018\footnote{
https://www.budget.gc.ca/2018/docs/themes/progress-progres-en.html?utm\_source=CanCa\&utm\_medium=Activities\_e\&utm\_content= Progress\&utm\_campaign=CAbdgt18}, 
the Canadian Government provided funding in support of a Digital Research Infrastructure Strategy to deliver more open and equitable access to advanced computing and big data resources for researchers across Canada.
The National Research Council’s (NRC’s) world-class scientists are an important part of this strategy, helping to advance Canadian academia and industry through cutting-edge innovation and reinforcing Canada’s research capabilities and strengths. Their facilities, expertise and networks help convene strategic, large-scale national teams committed to innovation. Budget 2018 helped to lower the cost of partnering with the NRC so more small and medium-sized enterprises, colleges and universities have been able to make use of its services. 
This plays out in terms of ML applications in astronomy through NRC-Herzberg, particularly through the new hires and strategic investments through the Canadian Astronomical Data Centre (CADC) and development of the Canadian Advanced Network For Astronomy Research.

{\bf NRC-Herzberg Astronomy \& Astrophysics (CADC/CANFAR):}  Current expertise in ML applications in Canadian astronomy are centred at or include significant collaboration with NRC-HAA.   Through CANFAR (operated by NRC's CADC), astronomers in Canada have direct and easy access to cyber infrastructure, growing numbers of GPUs, fast access to resources, fast and interactive processing, and safe storage.   These are necessary components for a competitive research program in the era of big data to remain competitive with the world.  The data that CADC has been archiving for over a decade also provides long baselines for validating new ML data methods, and providing testing data sets.  Furthermore, NRC-HAA CADC are a working model of the value and innovation possible when expertise is brought together from computer science, engineering, data science, etc. and focused on astronomy.

{\bf AI Research Institutions (MILA, Vector, AMII):} An equally important resource and new opportunity in Canada for the development of ML applications in astronomy are through our world-leading AI research centers: the Montreal Institute for Learning Algorithms (MILA, Montreal). the Vector Institute for Artificial Intelligence (Vector, Toronto), and the Alberta Machine Intelligence Institute (AMII, Edmonton).  These are not-for-profit research and educational institutions, closely partnered with the Universities, and partially supported through government funding. MILA and Vector were founded in 2017, while AMII has a 15 year history and only recently begun to specialize in ML applications.  All of these institutions work with industry, start-ups, incubators, and accelerators to advance AI research and drive its application, adoption, and commercialization across Canada.
Increasing our links with these specialized institutions, which are recognized as the world leaders in ML/AI, can provide Canadian astronomers with a competitive advantage in the era of big data science.

{\bf New Methodologies:} ML is a different way of thinking about computing problems and big data compared with traditional methods.  It goes beyond feature engineering to improve performance (e.g., detecting anomalies or reducing data uncertainties), and we will need to move beyond simply substituting traditional legacy programs with ML applications.  Some examples include astronomers at UVic and SFU who are exploring ML for detection of new astronomical phenomena \citep[e.g.,][]{Teimoorinia2016}, and even as a tool to learn new physics. This kind of innovative thinking is ideal for the ML research centres and their innovative think-tank methodologies, but combined with astronomers who know the motivating science questions.  ML gives us the opportunity to imagine using ML methodologies to e.g., control observatories, synchronize observing facilities so that there is more collaboration in observational planning, and even collate multi-wavelength datasets (imaging, spectroscopy, wavelength regions, spatial and wavelength resolution, time domain, polarization) in ways we have yet to imagine.

{\bf Advantages for Canadian Astronomical Research:}  Currently, there are several groups in Canada leading the development of ML applications for astronomical research. In Montreal, astronomers at McGill and UdM are working with researchers at MILA on ML applications for the analysis of Euclid data to study cosmological structure \citep{Hezaveh2019}.  In Victoria, several groups of astronomers are working with researchers at NRC on ML applications ranging from extragalactic studies \citep{Bottrell2019} to stellar spectroscopic surveys \citep{Bialek2019}.
In Toronto, several groups of astronomers are using ML techniques in both data science analyses \citep{leung2019}, and looking into ML for wavefront reconstruction and predictions for use in adaptive optics \citep[e.g., GIRMOS; ][]{Swanson2018}.
ML could even be embedded in the next generation of Canadian astronomical instrumentation (e.g., TMT-IRMOS) and surveys (MSE).  It also provides excellent opportunities for collaborations with industry (autonomous telescopes are alot less dangerous than autonomous vehicles), and provide outstanding opportunities for knowledge transfer and contributions to the Canadian economy. Finally, this new technique is known to draw top students and researchers from interdisciplinary fields, making it an ideal training tool for skilled researchers in the future.

\section{Summary} 
The growth of machine learning applications in astronomical research has been remarkable and powerful.   This new data analysis technique requires us to think differently about astronomical problems, develop new approaches to data science, and collaborate extensively with researchers in computer science, engineering, and other fields. It attracts some of the top HQP from around the world, and can foster training of a new generation of highly skilled researchers in astronomy and beyond.  The methodologies are highly transferrable and astronomy provides some of the largest, homogeneous data sets for application testing and development.  The knowledge transfer is extraordinary such that ML developments in astronomy can be directly related to industrial innovation and the Canadian economy, fields that are being lead by expertise at the Canadian ML research institutes (MILA, Vector, and AMII).   For Canadian astronomy itself, 
ML could be embedded in the next generation of astronomical instrumentation and surveys, such that ML pipelines and scientific analysis tools make it possible for science to come directly from the telescope.




\section {White Paper Criteria Section} 

\begin{lrptextbox}[How does the proposed initiative result in fundamental or transformational advances in our understanding of the Universe?]

As CMB observations have become increasingly more precise, inflation and the simplest $\Lambda$CDM models continue to be a successful explanation of the evolution of the universe.
However, the nature of the key components of these models remains unknown. Discovering the physical nature of the field(s) driving inflation, the source of the apparent accelerated expansion of the universe (dark energy), and the particle(s) constituting dark matter are the primary goals of modern cosmology. 

\textit{ML as a component of future large scale surveys:}
In the coming decade, a new generation of large surveys and telescopes (e.g., Euclid, LSST, CHIME) promise to transform cosmology as we know it. The large volumes of data produced by these instruments will allow significant improvements in the precision measurement of cosmological parameters, potentially allowing us to pinpoint specific models of dark energy and particle properties of dark matter, while opening new windows into the exploration of the physics of the universe. 
The analysis of these data requires new computational and statistical methods, not only for the best possible classification schemes but also for scientific inference.

\textit{ML integration with physical simulations:}
Complex astrophysical processes, combined with instrumental simulations face a computational bottleneck. The interplay between ML and physical models will continue to generate more creative ways to accelerate simulations, incorporate domain knowledge in ML models to be more data efficient, and better match between simulations and real data by learning how to fill the synthetic gap. 
The interpretability of ML methods is a new, active area of research \citep[e.g.,][]{2017arXivDishi-Velez}, and astronomy-specific applications in which we already have physical understanding could provide interesting test cases for interpretability studies.

\textit{ML as tool for discovery:} In the medical field, ML has revealed previously unknown gender differences in retinal images, for example \citep{Poplin2018}. Therefore, ML has tremendous potential as a direct vehicle for discovery.
Astronomers are only beginning to explore the power of ML to answer unsolved mysteries, making ML an exciting frontier field.
New ML techniques, expanding software eco-system, and ML specialized hardware will permit drastic improvements in efficiencies in the analysis of complex simulations and large data sets.
ML is expected to emerge as a necessary tool to fully exploit the vast majority of survey data in the coming decade. 
In this sense, ML will not only be an asset, but a necessity.

\end{lrptextbox}

\begin{lrptextbox}[What are the main scientific risks and how will they be mitigated?]

One of the main critiques of ML is that the algorithms are a black box, resulting in uncertain interpretations and poor/misguided errors analyses. These risks can be mitigated through comparisons with traditional methods (which typically have a different set of uncertainties), benchmark data sets, incorporating physics into the ML models, and statistical methods.  Additional tools, such as the generation of saliency maps, have been developed by the larger ML community to aid in interpretation.
Thus, the risks are not specific to astrophysics, and often develop into interesting and active research topics, such as adversarial or interpretable ML with promising results.  
While ML associated risks can be mitigated, training our community in best practices and collaborating with ML researchers will be essential to ensuring that these techniques are used correctly. 

There is evidence that the Canadian astronomy community wants more focused training in statistics (see the LRP2020 white paper \emph{Astrostatistics in Canada}), and a desire for training in ML techniques likely exists as well. Collaborating with ML researchers and with statisticians will improve our ability to interpret ML outputs.
Recent development in deep learning tend to show larger ML models with larger data sets perform better, and it could transform into a risk for Canadian scientific institutions if we do not have access to the large infrastructure needed to train such models. Mitigating this risk would involve co-development with ML-research in resource efficient and high performance methods for astrophysical research, while also securing access to ML friendly digital infrastructure.

\end{lrptextbox}

\begin{lrptextbox}[Is there the expectation of and capacity for Canadian scientific, technical or strategic leadership?] 

Canada is a world leader in the development of the field of machine learning, attracting large investments and skilled researchers to its prestigious AI Research Institutes (MILA, Vector, AMII). This provides unique opportunities for Canada to also be a leader in the application of ML in the field of astrophysics. 

\end{lrptextbox}

\begin{lrptextbox}[Is there support from, involvement from, and coordination within the relevant Canadian community and more broadly?] 
Established Canadian astronomers associated with large scale surveys (McGill, UdM, UVic, UBC, Toronto, UWO, etc.) directly benefit from ML. The Canadian ML-astro community is growing with a few recent hires (NRC, UdM). The Canadian astronomical instrumentation community is also starting to look into ML applications for adaptive optics, RFI filtering, image processing. One could easily imagine ML being introduced into observatory operations, such as queue scheduling, remote observing, and inter-observatory operations.
The recent wave of new cross-disciplinary institutes at Canadian universities (McData at McMaster, Matrix at UVic, UBC Data Science Institute) have attracted funds and students, and the astrophysics community should seek more participation, driven by our ML interests and involvement in big data surveys.
Altogether the Canadian community is involved in various ML activities but could benefit from more unification through organised meetings, collaboration with Canadian ML hubs, the Canadian ML Research Institutes, and industry participation.

\end{lrptextbox}

\begin{lrptextbox}[Will this program position Canadian astronomy for future opportunities and returns in 2020-2030 or beyond 2030?] 

Machine learning methods have seen a rapid expansion and breakthroughs in the past few years, often lead by the Canadian AI Research Institutes (MILA, Vector, and AMII). The simplicity of implementing ML models, their remarkable power in finding complex patterns, and their adaptability to many different problems has resulted in their widespread use in different fields. 
Astronomy provides new clean and robust applications to assess their efficacy, as well as new challenges.  Conversely, they provide astronomers new tools for improving speed and precision in data analysis.  For two main reasons, these methods could transcend current methods in astrophysics and cosmology, putting Canadian astronomy at an extreme advantage in the 2020's and beyond.


Simultaneously, the data rates for upcoming larger facilities, like SKA, will be another factor of 10-100x higher. ML has two distinct advantages; (1) speed and automation, and (2) deep learning networks can learn complex, high-order, non-Gaussian priors from their training data.  These together can result in higher precision and accuracy for many astronomical problems.
By ramping up now in the 2020's, we can expect Canadian astronomy to be in a leadership position throughout the 2030's. 


 
\end{lrptextbox}

\begin{lrptextbox}[In what ways is the cost-benefit ratio, including existing investments and future operating costs, favourable?] 

Costs are currently the scientific choices of Canadian astronomy faculty, and commitments made at NRC-CADC.   But we advocate for infrastructure that can keep up with these needs.  Currently Compute Canada is not able to keep pace with demand, with limited storage capacity and a short-term per user/per group accounts.   Also, ML requires some overhead for users, thus modules that make it easier and quicker for users to access ML benefits should be developed.   We suggest that this should be done in coordination with the ML leaders in Canada (e.g., MILA, Vector, AMII), as experts in computational techniques and statistical methodologies for big data.

\end{lrptextbox}

\begin{lrptextbox}[What are the main programmatic risks and how will they be mitigated?]

\textit{Technical readiness:}
The research pace in ML has been extremely rapid in the past few years, and it has been difficult even for ML researchers to keep up. While it is not clear that this pace will continue, a lag in the astrophysical implementations could mean losing our current competitive edge.  To avoid that loss requires an increase in human resources (e.g., accessible user modules and computational help for applications), but also an increase in infrastructures (e.g., fast computers and processors, large and safe storage, easy access to and interactions with computing and data resources).

\noindent
\textit{Governance plan:}
It is not yet clear where the raw data from the next generation of large surveys will be stored, but even working on the reduced data sets will be larger than anything we are yet used to.   
If storage is not in Canada, then we may only have the facilities to store processed data, made more complex by issues related to data rights.   For spectroscopy (SDSS, PFS, MSE), then raw data and reduced spectra will likely be held by the survey institutions, but users should be able to download entire spectral libraries and parameter tables.   While this is not as computing intensive, these can still be large files and require significant storage space.
Modules to help users get the benefits of ML will need to be centralized (e.g. at NRC-CADC), and therefore we will need a platform for sharing these across all Canadian Universities.  
Thus, we will need to develop a governance model around safe, reliable, and large data storage for Canadian astronomers to  leverage our current investments and reach our scientific goals.

\end{lrptextbox}

\begin{lrptextbox}[Does the proposed initiative offer specific tangible benefits to Canadians, including but not limited to interdisciplinary research, industry opportunities, HQP training, EDI, outreach or education?] 

\textit{Interdisciplinary and industrial opportunities:}
ML is interdisciplinary by nature, being lead by computer science and statistics, and involving engineering, scientists, medical researchers.   The benefits of investing in ML to Canadians are wide spread, ranging from technological transfers with research into self-driving cars, natural language processing, fraud detection, robot navigation, medical diagnostics, targeted marketing, and gaming, to knowledge transfer opportunities as students and researchers work on both scientific/astronomical problems that then be applied to industry applications.  Because ML skills and training are in high demand and highly transferrable, the field is very attractive to new students.  This means some of the top young researchers will at attracted to astronomical problems, providing our Canadian astronomical community with outstanding researchers.   

\noindent
\textit{Equity, Diversity, and Inclusivity:}  ML is attracting women and researchers from other under-represented groups.  We note that nearly 50\% of the co-authors on this white paper are women, and most of us have (co-)supervised women students on ML related projects.  Furthermore, our ML applications and big data requirements can provide outstanding research and employment opportunities for young Canadians and attract top international researchers to Canada.  One group whom we need to reach out to more with respect to astronomical research in general, and ML specifically, are the indigenous Canadian communities.   

\noindent
\textit{Outreach and education:} Astronomy has always captured the public's imagination like no other science.  The public is already familiar with ML in the form of image searches and facial recognition on Google and Facebook, and will be naturally fascinated by the use of ML in astronomy.  This initiative will have the potential to inspire schools and young people to learn programming skills early.

\end{lrptextbox}

\bibliography{mlwp} 

\end{document}